\documentclass[a4paper,amsmath,amssymb]{jpconf}
\usepackage{graphicx}%
\usepackage{amsmath, amsthm, amssymb}
\usepackage{latexsym}
\usepackage{psfrag}

\begin{document}

\def\bea{\begin{eqnarray}}
\def\eea{\end{eqnarray}}
\def\beq{\begin{equation}}
\def\eeq{\end{equation}}
\def\f{\frac}
\def\k{\kappa}
\def\e{\epsilon}
\def\ve{\varepsilon}
\def\be{\beta}
\def\D{\Delta}
\def\h{\theta}
\def\t{\tau}
\def\a{\alpha}

\def\rk{\rho^{ (k) }}
\def\rek{\rho^{ (1) }}
\def\cek{C^{ (1) }}
\def\rz{\bar\rho}
\def\rt{\rho^{ (2) }}
\def\rtb{\bar \rho^{ (2) }}
\def\trk{\tilde\rho^{ (k) }}
\def\trek{\tilde\rho^{ (1) }}
\def\trz{\tilde\rho^{ (0) }}
\def\trt{\tilde\rho^{ (2) }}
\def\r{\rho}
\def\tD{\tilde {D}}

\def\s{\sigma}
\def\kb{k_B}
\def\la{\langle}
\def\ra{\rangle}
\def\nn{\nonumber}
\def\up{\uparrow}
\def\dn{\downarrow}
\def\S{\Sigma}
\def\dg{\dagger}
\def\d{\delta}
\def\p{\partial}
\def\l{\lambda}
\def\L{\Lambda}
\def\G{\Gamma}
\def\o{\Omega}
\def\w{\omega}
\def\g{\gamma}

\def\noi{\noindent}
\def\a{\alpha}
\def\d{\delta}
\def\p{\partial} 

\def\la{\langle}
\def\ra{\rangle}
\def\e{\epsilon}
\def\n{\eta}
\def\g{\gamma}
\def\break#1{\pagebreak \vspace*{#1}}
\def\hf{\frac{1}{2}}

\title{Stochastic models of classical particle pumps : Density dependence of directed current}
%\title{Density dependence of current in exclusion process of pumped particles} %: Presence and absence of current reversal}
\author{Debasish Chaudhuri}
\address{
Indian Institute of Technology Hyderabad,
Yeddumailaram 502205, Telengana, India
}

\ead{debc@iith.ac.in}

\date{\today}

\begin{abstract}
We present and compare different versions of a simple particle pump-model that describes average directed current of repulsively interacting particles in a narrow channel, due to time-varying local potentials. We analyze the model on discrete lattice with particle exclusion, using three choices of potential-dependent hopping rates that obey microscopic reversibility. Treating the strength of the external potential as a small parameter with respect to thermal energy, we present  a perturbative calculation to obtain the expression for average directed current. This depends on driving frequency, phase, and particle density. The directed current  vanishes as density goes to zero or close packing. For two choices of hopping rates, it reaches maximum at intermediate densities, while for a third choice, it shows a curious current reversal with increasing density. This can be interpreted in terms of a particle-hole symmetry. Stochastic simulations of the model show good agreement with our analytic predictions.
\end{abstract}

\section{Introduction}
In everyday life we often use mechanical devices that utilize oscillatory force along with a valve mechanism to generate unidirectional flow, e.g., a simple hand pump. In this, the valve breaks space-inversion symmetry, as a result breaking time-reversal symmetry, leading to unidirectional flow.
The same basic principle of oscillatory forcing, and time-reversal symmetry breaking leading to directed motion, is utilized in operation of various molecular motors~\cite{Julicher1997, Reimann2002, Astumian2002}, and ion pumps~\cite{Gadsby2009, Astumian2003} in biological cell, and to generate overall unidirectional flow of electrons in quantum pumps~\cite{Brouwer1998, Sela2006,Cavaliere2009, Strass2005}. However, in all these cases noise, stochastic or quantum, plays important role in the resultant dynamics. In molecular motors, e.g., repeated hydrolysis of ATP leads to a stochastic oscillatory energy input, and the intrinsic head-tail directionality of polymeric track on which the motors move, breaks the inversion symmetry to act like a valve~\cite{Julicher1997}.  
Most theoretical studies of stochastic pumps have discussed properties of noninteracting system of particles, 
apart from a few exceptions~\cite{Derenyi1995,Derenyi1996,Aghababaie1999,savel2004}.

In this paper, we present a stochastic particle pump model in which an external time-varying potential pumps energy, and a spatially varying phase factor of the oscillation breaks time-reversal symmetry to generate a directed current. We particularly focus on the effect of inter-particle interaction on the dynamics.
This interaction may be incorporated via an exclusion process in a spatially discretized version of the dynamics. Two variants of this model have already been  proposed and analyzed in some detail~\cite{Jain2007, Marathe2008,Chaudhuri2011}. 
Here we present a unified description of the model on discrete lattice. The model allows for several choices of hopping rates dependent on instantaneous local potential, where each choice obeys microscopic reversibility. We show that depending on this choice, one obtains different forms of density dependence of average directed current.

\section{Model}
We consider particles hopping on a ring, discretizing space into $s=1,\dots, N$ lattice sites, such that the system size is $L=N$ in units of  lattice spacing $b$.  
We assume that particles evolve under a position dependent weak oscillatory potential 
$\be V_s = \l_s \sin(\o t + \phi_s)$ where $\be=1/\kb T$ with Boltzmann constant $\kb$ and temperature $T$, and $\phi_s$ denote local phase factor. %such that $\l_s \ll 1$ 
This potential drives the system out of equilibrium. 
If the driving frequency $\o$ is slow with respect to the diffusion time scale $1/f_0 = a^2/D$ where $D$ denotes the diffusivity, 
the system of particles would come to local thermal equilibrium with the instantaneous local potential. We assume microscopic reversibility, 
i.e., the hopping rates are such that given the value of local potential at any instant of time the detailed balance condition,
$ n_s w_{s, s+1}=  n_{s+1}  w_{s+1,s}$, is obeyed.  In this relation $n_s$ stands for the occupation number of $s$-th site and $w_{s, s\pm 1}$ is the time-dependent hopping rate from $s$-th to $s\pm 1$-th site.  At each moment the system tries to reach equilibrium distribution corresponding to the instantaneous potential, but lags behind as the potential itself changes with time.  This keeps the system out of equilibrium.  Following three choices of particle hopping rates $w_{s,s\pm 1}$ obey local detailed balance:

(A) $w_{s, s \pm 1} = f_0 \exp[\be V_s]$, a symmetric hopping rate that depends only on the on-site potential energy~\cite{Jain2007,Marathe2008};  

(B) $w_{s, s \pm 1} = f_0 \exp[-\be(V_{s \pm 1}-V_s)/2]$, depends on relative strength of the potential energies~\cite{Chaudhuri2011}; 

(C) $w_{s, s \pm 1} = f_0 \exp[-\be V_{s \pm 1}]$ depends only on the potential energy at the site where the particle hops to.  

In this paper we present a {\em unified derivation} of the time averaged DC current obtained for these three cases. While models B and C are able to pump DC current even 
in a non-interacting system of particles, in model A  interaction is crucial in order to achieve pumping. 

We present analytic results using a perturbation theory proposed in Ref.~\cite{Chaudhuri2011}, and compare them with numerical simulations. We present a detailed comparative analysis of the three models of $w_{s, s \pm 1}$  proposed above, two of which (models A and B) were already discussed earlier~\cite{Marathe2008,Chaudhuri2011}, and the third one (model C) being the main new result of this paper.

A hard core repulsion between particles is modeled by exclusion process, in which two particles can not occupy the same lattice site. With this restriction, the local density $\r_s=\la n_s \ra$ and two-point correlation functions $C_{s,p} = \la n_s n_p \ra$ obey the following dynamics,
\bea
\f{d \la n_s\ra}{dt} &=& w_{s-1,s} \la n_{s-1}(1-n_s)\ra +w_{s+1,s}\la n_{s+1}(1-n_s)\ra \crcr
&-&  w_{s,s-1}\la n_s(1-n_{s-1})\ra - w_{s,s+1}\la n_s(1-n_{s+1})\ra.\\
\f{d \la n_s n_p\ra}{dt} &=& 
w_{s-1,s} \la n_{s-1}(1-n_s) n_p\ra +w_{s+1,s}\la n_{s+1}(1-n_s) n_p\ra \crcr
&+& w_{p-1,p} \la n_s n_{p-1}(1-n_p)\ra +w_{p+1,p}\la n_s n_{p+1}(1-n_p)\ra \crcr
&-&  w_{s,s-1}\la n_s(1-n_{s-1}) n_p\ra - w_{s,s+1}\la n_s(1-n_{s+1}) n_p\ra \crcr
&-&  w_{p,p-1}\la n_s n_p(1-n_{p-1})\ra - w_{p,p+1}\la n_s n_p(1-n_{p+1})\ra \\
\f{d \la n_s n_{s+1}\ra}{dt} &=& 
w_{s-1,s} \la n_{s-1}(1-n_s) n_{s+1}\ra  +w_{s+2,s+1}\la n_s n_{s+2}(1-n_{s+1})\ra \crcr
&-&  w_{s,s-1}\la n_s(1-n_{s-1}) n_{s+1}\ra - w_{s+1,s+2}\la n_s n_{s+1}(1-n_{s+2})\ra
\label{eq:motion}
\eea
The last equation is for the special case of nearest neighbor correlations.
Writing $\r_s = \la n_s \ra$ and the multi-point correlations as $C_{s,p,\dots} = \la n_s n_p \dots \ra$
we can re-express the above relations as
\bea
\f{d \r_s}{dt} &=& w_{s-1,s} (\r_{s-1}-C_{s-1,s}) + w_{s+1,s} (\r_{s+1}-C_{s,s+1}) \crcr
&-&  w_{s,s-1}(\r_s- C_{s-1,s}) - w_{s,s+1} (\r_s- C_{s,s+1}).\\
\f{d C_{s,p}}{dt} &=& 
w_{s-1,s} (C_{s-1,p} - C_{s-1,s,j}) +w_{s+1,s} (C_{s+1,p}-C_{s, s+1, p}) \crcr
&+& w_{p-1,p} (C_{s,p-1}-C_{s,p-1,p}) +w_{p+1,p} (C_{s,p+1}-C_{s,p,p+1}) \crcr
&-&  w_{s,s-1} (C_{s,p} - C_{s-1,s,p}) - w_{s,s+1} (C_{s,p}-C_{s,p,p+1}) \crcr
&-&  w_{p,p-1} (C_{s,p} - C_{s,p-1,p}) - w_{p,p+1} (C_{s,p}-C_{s,p,p+1}) \\
\f{d C_{s,s+1}}{dt} &=& 
w_{s-1,s} (C_{s-1,s+1}-C_{s-1,s,s+1})  +w_{s+2,s+1} (C_{s,s+2} - C_{s,s+1,s+2}) \crcr
&-&  w_{s,s-1} (C_{s,s+1} - C_{s,s-1,s+1})  - w_{s+1,s+2} (C_{s,s+1} - C_{s,s+1,s+2}). 
\eea
Thus dynamics of each order of correlation depends on correlations of higher order, following a Bogoliubov-Born-Green-Kirkwood-Yvon (BBGKY) hierarchy. 
Note that the evolution of local density may be represented as $d\r_s / dt = J_{s-1,s} - J_{s,s+1}$ where the local current
\beq
J_{s-1,s} =  (w_{s-1,s} \r_{s-1} - w_{s,s-1} \r_s) - (w_{s-1,s} - w_{s,s-1} ) C_{s-1,s} 
\eeq 
In the time periodic steady state, local current averaged over the period $\t = 2\pi / \o$ is independent of position. Therefore the net time- and space- averaged 
directed current is given by 
\beq
\bar J = \f{1}{N \t} \sum_{s=1}^N \int_0^\t dt J_{s-1,s}.
\eeq

For potential strength $\l_s=0$ at all lattice sites, the above model reduces to homogeneous symmetric
exclusion process, characterized by the following local density and correlation
functions
\bea
\rz &=&  \f{n}{L} =\r, \crcr
C^{(2)} &=&  \r \f{n-1}{L-1} \crcr
C^{(3)} &=&  C^{(2)} \f{n-2}{L-2} 
\eea
etc.~\cite{Schutz2000}, where $n$ is the total number of particles. As we show in the following, the BBGKY hierarchy separates by order, if one expands local quantities like $\r_s$, $C_{s,p}$ etc. in perturbative expansion around $\l_s=0$.
This allows one to obtain exact expressions within perturbative expansion.

\section{Perturbative calculation}

We consider driving at all the sites with constant potential strength and frequency
\bea
\be V_s = \l \sin(\o t +\phi_s) = \l\times u_s
\eea
where 
\bea
u_s &=& 2\mbox{Re}~ [\n_s e^{i\o t}], \crcr
\n_s &=& -\f{i}{2} e^{i \phi_s}.
\eea

For small values of $\l$ we can linearize the transition rates to obtain :
(A) $w_{s, s\pm 1} = f_0 (1+ \l u_s)$,  
(B) $w_{s, s\pm 1} = f_0 \{1-\hf \l(u_{s\pm 1}-u_s) \}$  
(C) $w_{s, s\pm 1} 
= f_0 (1- \l u_{s\pm 1})$ 
and the corresponding bond currents are expressed as
\begin{description}
\item[(A) $J_{s-1,s} = -f_0 (\r_s - \r_{s-1} ) - \l f_0  (u_s \r_s - u_{s-1} \r_{s-1} ) + \l f_0  (u_s - u_{s-1}) C_{s-1,s}$,  ] 
\item[(B) $J_{s-1,s} = -f_0 (\r_s - \r_{s-1} ) - (\l f_0/2) (u_s - u_{s-1} )(\r_{s-1} + \r_s - 2 C_{s-1,s})$  ]
\item[ (C) $J_{s-1,s} = -f_0 (\r_s - \r_{s-1} ) - \l f_0 (u_s \r_{s-1} - u_{s-1} \r_{s} ) + \l f_0  (u_s - u_{s-1}) C_{s-1,s}$. ] %,  $p=s \pm 1$.  ]
\end{description}

Therefore the space-time averaged directed current is
\bea
\bar J_A &=& \f{\l f_0}{N\t} \sum_{s=1}^N \int_0^\t dt \,(u_s-u_{s-1})C_{s-1,s}  \nn\\
\bar J_B &=& -\f{\l f_0}{2 N \t} \sum_{s=1}^N \int_0^\t dt \, (u_s-u_{s-1})(\r_{s-1}+\r_s - 2 C_{s-1,s}) \nn \\
\bar J_C &=& -\f{\l f_0}{ N\t} \sum_{s=1}^N \int_0^\t dt \, [ (u_s \r_{s-1} - u_{s-1} \r_{s} ) - (u_s - u_{s-1}) C_{s-1,s}] 
\label{eq:JC}
\eea
Note that, for non-interacting particles the correlation function $C_{s-1,s} =0$ leads to $\bar J_A = 0$. That means, within model-A, the pump will
drive particles in an averaged unidirectional fashion only in the presence of interaction -- free particles can not be pumped within this model~\cite{Marathe2008}. However, for the
other two models this requirement is absent. Even free particles may be pumped in a unidirectional manner.

Let us write down the density and correlation functions as  perturbative expansion in potential strength $\l \,(\ll 1)$
\bea
\r_s &=& \r + \sum_{k=1,2,\dots} \l^k \rk_s \crcr
C_{s,p} &=& C^{(2)} + \sum_{k=1,2,\dots} \l^k C_{s,p}^{(k)}.
\label{perturb}
\eea

\begin{figure}[t] 
\begin{center}
\includegraphics[width=7.9 cm] {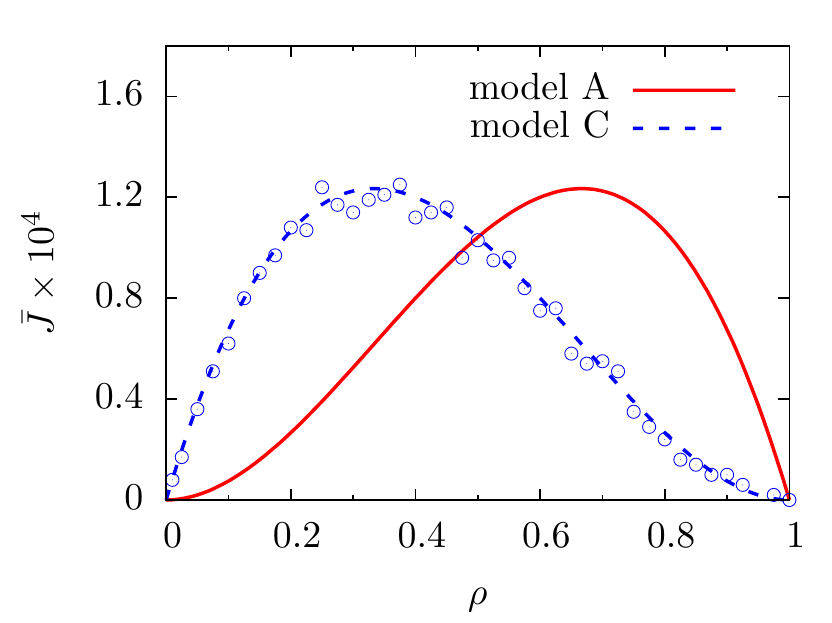} 
\includegraphics[width=7.9 cm] {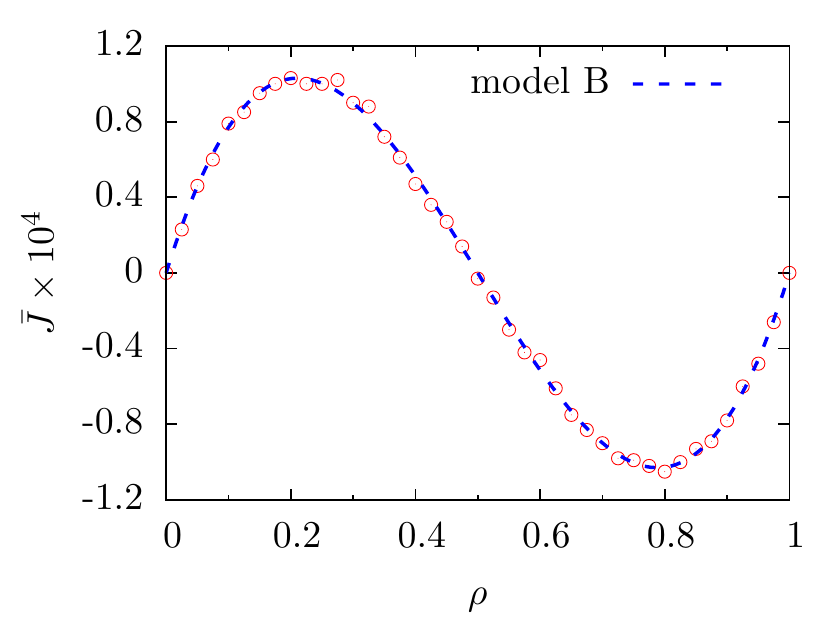} 
\caption{(Color online) 
Directed current $\bar J$ as a function of mean density 
$\rz$. The points denote Monte-Carlo result  and the lines are plot of the functions in  Eq.\ref{eq:JC1}.
In plotting the functions we replaced $q_0$ by $\r(1-\r)$ and $k_0$ by $\r^2(1-\r)$.
Left panel: Shows plots of analytic functions for models A and C, and simulation data for model C. 
Right panel: Comparison of simulation results of model B with analytic prediction. 
The parameters used are system size $L=16$, bare hopping rate due to free particle diffusion $f_0 = 0.34$, potential strength $\l=0.5$, frequency of
oscillation $\o = 0.2 \pi$ such that the time period $\t = 10$. We choose phase difference between consecutive lattice points $\phi = \pi/2$. The data (points) were collected over $100\t$ after 
equilibration over $100\t$. All the data for model C were averaged over $10^5$ initial conditions, and for model B over $10^6$ initial conditions.
}
\label{fig:Jro}
\end{center}
\end{figure}

We consider the case where the phase factor of driving potential $\phi_s = \phi s$ where $\phi=2\pi/L$, remembering $L$ is expressed in units of lattice parameter. 
Within the perturbative expansion, the above relations for directed current may be calculated exactly~\cite{Marathe2008, Chaudhuri2011}. The results for the
first two cases were derived earlier, and the third one is presented in this paper. In what follows we derive all three results. However, before we start deriving 
them, let us enlist the expressions for current corresponding to the three variants of the model here,
\bea
\bar J_A &=& -2 \l^2 k_0 f_0^2 \f{ \o \sin \phi (1-\cos \phi)}{\o^2 + 4 f_0^2 (1-\cos \phi)^2} \nn\\
\bar J_B &=& \l^2  (q_0 - 2 k_0) f_0^2 \f{\o \sin \phi (1-\cos \phi)}{\o^2 + 4 f_0^2 (1-\cos \phi)^2} \nn\\
\bar J_C  &=&  2\l^2  (q_0 - k_0) f_0^2 \f{\o \sin\phi (1-\cos\phi)}{\o^2 + 4f_0^2 (1-\cos\phi)^2 }, 
\label{eq:JC1}
\eea
with $q_0 = \r - C^{(2)}$, $k_0 = C^{(2)} - C^{(3)}$. 
Note that in the limit of large $n$ and $L$ keeping $\r = n/L$ constant, $q_0 \approx \r (1-\r) $ and $k_0 \approx \r^2(1-\r)$. 
Thus $\bar J_A \sim \r^2 (1-\r)$, and $\bar J_B \sim \r(1-\r)(1-2\r)$, and $\bar J_C \sim \r(1-\r)^2$ (See Fig.\ref{fig:Jro}). 

%%%%%%%%%%%%
\subsection{Time evolution and solution}

Let us first consider model C, which constitutes the main new contribution of this paper. Using the perturbative expansion of Eq.(\ref{perturb}),  time evolution of the first order perturbations can be written as,
\bea
\f{d\rek_s}{dt} &=& f_0 \D_s \rek_s + f_0 q_0 \D_s u_s,   
\label{eq:rek} \\
\f{d \cek_{s,p}}{dt} &=& f_0 (\D_s + \D_p)\cek_{s,p} + f_0 k_0 (\D_s u_s+ \D_p u_p) ~~ {\rm for~}  p\neq s \pm 1, \crcr
\f{d \cek_{s,s+1}}{dt} &=& f_0 (\cek_{s-1,s+1} + \cek_{s,s+2} -2 \cek_{s,s+1} )+ f_0 k_0 (u_{s-1} + u_{s+2}  - u_{s} -u_{s+1})
\label{eq:cek}
\eea
where $\D_s g_{s,p} = g_{s+1,p} + g_{s-1,p} - 2 g_{s,p}$. 
Note that the above time evolution, for
first order terms in perturbative expansion, remains the same for all the three variants of the model considered above. This is easy to see by comparing with 
Ref.s~\cite{Marathe2008,Chaudhuri2011}.

These linear differential
equations can be solved exactly to find long time limit of time-varying steady state~\cite{Chaudhuri2011},
\bea
\rek_s(t) = 2 {\rm Re} [A_s^{(1)} \exp(i \o t)].
\eea
Using this in Eq.(\ref{eq:rek}) we find, for all $s$,
\beq
-A_{s-1}^{(1)} - A_{s+1}^{(1)} + (2 + i \o/f_0) A_s^{(1)} = - q_0 (-\eta_{s+1} - \eta_{s-1}+2\eta_s). 
\eeq
Clearly this equation  can be written in the operator form,
\beq
\hat Z \mid A\ra = -q_0 \hat \D \mid \eta\ra
\eeq
with matrix elements
\bea
\D_{s,p} &=&  \d_{s,p+1}  +\d_{s,p-1} -  2  \d_{s,p}. \crcr
Z_{s,p} &=& \D_{s,p} - \f{i \o}{f_0}\d_{s,p} %=  -\d_{s,p+1} + (2 + i \o/f_0) \d_{s,p} - \d_{s,p-1} 
\eea
where $\hat\D$ satisfies the eigenvalue equation $\hat \D | q \ra = \e_q | q \ra$ with $\e_q = -2(1-\cos q)$
and the eigenfunction $\psi_s(q) = \la s | q\ra = (1/\sqrt L) \exp(-i q s)$ where $q=2\pi k/L = \phi\, k$ where $k=1,2,\dots,N$
such that $\psi_{s+N}=\psi_s$. Similarly, $\hat Z | q \ra = (\e_q - i\o/f_0)  | q \ra$. Thus the solution may be expressed as
$| A\ra = -q_0 \hat Z^{-1}  \hat \D | \eta\ra$.

In the real-space representation 
$\la s | A\ra = -q_0 \sum_{q,m}\la s| \hat Z^{-1} | q\ra \la q |  \hat \D | m\ra \la m | \eta\ra
= -q_0 \sum_{q,m} (\e_q - i\o/f_0)^{-1} \la s\mid q\ra \e_q \la q\mid m\ra \eta_m 
= -q_0  \sum_{q,m}   (\e_q - i\o/f_0)^{-1} \e_q \psi_s(q) \psi^\ast_m(q) \eta_m$. Therefore,
\bea
A_s^{(1)} &=& -q_0   \sum_{m=1}^N \sum_{k=1}^N \f{\e_{\phi k}}{\e_{\phi k} - i\o/f_0} \psi_s(\phi k) \psi^\ast_m(\phi k) \eta_m \crcr
&=& \f{i q_0}{2} e^{i\phi s}  \f{\e_{\phi}}{\e_{\phi} - i\o/f_0}, 
\eea
where $\e_\phi = -2 (1-\cos\phi)$.
The equation for two point correlation function can also be solved~\cite{Chaudhuri2011}
\beq
\cek_{s,p} (t) = \f{k_0}{q_0} [ \rek_s(t) + \rek_p(t)] = 2 {\rm Re} [ A^{(1)}_{s,p} e^{i \o t}]
\eeq
where $A^{(1)}_{s,p}  = (k_0/q_0) (A_s^{(1)}  + A_p^{(1)} )$.

\subsection{Averaged directed current}
Using the above relations, one can calculate the space-time averaged directed current in the time-periodic steady state for all three variants of the
model, through the relations shown in Eq.(\ref{eq:JC}). For the particular case of model C, using the last relation in Eq.(\ref{eq:JC}), we have
\bea
\bar J_C &=& - \f{\l^2 f_0}{N \t} \sum_{s=1}^N \int_0^\t dt \left[  \left(1-\f{k_0}{q_0} \right) (u_s \rek_{s-1} - u_{s-1}\rek_s) - \f{k_0}{q_0} (u_s \rek_s - u_{s-1}\rek_{s-1})\right] \crcr
&=& - \f{\l^2 f_0}{N } \sum_{s=1}^N  2\, {\rm Re}\left[  \left(1-\f{k_0}{q_0} \right) (\eta_s^\ast A^{(1)}_{s-1} - \eta^\ast_{s-1} A^{(1)}_s) - \f{k_0}{q_0} (\eta_s^\ast A^{(1)}_s - \eta^\ast_{s-1} A^{(1)}_{s-1})\right]
\label{JC2}
\eea
where in the last step we have used the fact that after integration over a period $\t=2\pi/\o$, only the time-independent combinations of $u_s \rek_s$ terms which can be expressed as 
$ 2\, {\rm Re} [\eta_s^\ast A_s^{(1)}]$ etc. remain non-zero.

Given that $\eta_s = (-i/2) e^{i\phi s}$, and as we may write $A_s^{(1)}  = (i q_0/2) e^{i\phi s} a$ where $a=\e_\phi/[\e_\phi - i\o/f_0]$, the terms in Eq.(\ref{JC2}) may be evaluated. We find that 
$\eta_s^\ast A^{(1)}_s = - q_0 a/4$, and $\eta^\ast_{s-1} A^{(1)}_{s-1} = - q_0 a/4$. Thus the second term in the parentheses $(\eta_s^\ast A^{(1)}_s - \eta^\ast_{s-1} A^{(1)}_{s-1})=0$. Similarly one can
show that $(\eta_s^\ast A^{(1)}_{s-1} - \eta^\ast_{s-1} A^{(1)}_s) = (q_0/2) \sin\phi\, (i a)$. Thus the terms $2 {\rm Re}(\eta_s^\ast A^{(1)}_{s-1} - \eta^\ast_{s-1} A^{(1)}_s) = q_0 \sin\phi\, {\rm Im}(a)$.
Therefore,
\bea
\bar J_C &=& - \l^2 f_0  \left(1-\f{k_0}{q_0}\right)  \,\,q_0 \sin\phi\, {\rm Im}(a) \crcr
&=&  -\l^2   (q_0 - k_0) f_0^2 \f{\o \sin\phi \e_\phi}{\o^2 + f_0^2 \e_\phi^2 } \crcr
&=&  2\l^2  (q_0 - k_0) f_0^2 \f{\o \sin\phi (1-\cos\phi)}{\o^2 + 4f_0^2 (1-\cos\phi)^2 },
\eea
where in the last step we used the expression for $\e_\phi$.
In the limit of large system size $q_0-k_0 \approx \r(1-\r)^2$, and thus $\bar J_C \sim  \r(1-\r)^2$.

Similar arguments may be used to derive the results corresponding to models A and B. For example using relations in Eq.(\ref{eq:JC}),
\bea
\bar J_A &=& \f{\l^2 f_0}{N\t } \sum_{s=1}^N \int_0^\t dt  \f{k_0}{q_0} (u_s - u_{s-1}) (\rek_s + \rek_{s-1}) \crcr
&=& \f{\l^2 f_0}{N }  \f{k_0}{q_0} \sum_{s=1}^N  2\, {\rm Re} \left[ (\eta_s^\ast - \eta^\ast_{s-1}) (A^{(1)}_s + A^{(1)}_{s-1} )\right].
\eea
One can show that $(\eta_s^\ast - \eta^\ast_{s-1}) (A^{(1)}_s + A^{(1)}_{s-1} ) = (q_0/2) \sin\phi\, (i a)$, and thus $2\, {\rm Re} \left[ (\eta_s^\ast - \eta^\ast_{s-1}) (A^{(1)}_s + A^{(1)}_{s-1} )\right] 
= q_0 \sin\phi\, {\rm Im}(a)$. Thus
\bea
\bar J_A  &=& \l^2 f_0  \f{k_0}{q_0} q_0 \sin\phi\, {\rm Im}(a) \crcr
&=& - 2\l^2 k_0 f_0^2 \f{\o \sin\phi (1-\cos\phi)}{\o^2 + 4f_0^2 (1-\cos\phi)^2 }. 
\eea

Again, using Eq.(\ref{eq:JC})
\bea
\bar J_B &=& - \f{\l^2 f_0}{2 N \t} \left(1-\f{2 k_0}{q_0} \right) \sum_{s=1}^N  \int_0^\t dt\, (u_s - u_{s-1}) (\rek_{s-1}+\rek_s) \crcr
&=& - \f{\l^2 f_0}{2N} \left(1-\f{2 k_0}{q_0} \right) \sum_{s=1}^N 2\, {\rm Re} \left[ (\eta_s^\ast - \eta_{s-1}^\ast) (A_{s-1}^{(1)} + A_s^{(1)}) \right].
\eea
The relation $(\eta_s^\ast - \eta_{s-1}^\ast) (A_{s-1}^{(1)} + A_s^{(1)}) = (q_0/2) \sin\phi\, (i a)$, leading to $ 2\, {\rm Re} \left[ (\eta_s^\ast - \eta_{s-1}^\ast) (A_{s-1}^{(1)} + A_s^{(1)}) \right] 
= q_0 \sin\phi\, {\rm Im}(a)$. Thus one gets
\bea
\bar J_B &=& - \f{\l^2 f_0}{2} \left(1-\f{2 k_0}{q_0} \right) \, q_0 \sin\phi\, {\rm Im}(a) \crcr
&=& \l^2 (q_0 - 2 k_0) f_0^2 \f{\o \sin\phi (1-\cos\phi)}{\o^2 + 4f_0^2 (1-\cos\phi)^2 }. 
\eea

\section{Simulation}

A detailed numerical simulation for model A was presented earlier in Ref.~\cite{Jain2007}.
Here we perform Monte-Carlo simulations of the models B and C and present density dependence of directed current $\bar J$ in Fig.~\ref{fig:Jro}. In the stochastic simulation,
we randomly choose a lattice site $s$ with uniform probability and perform a trial move with rate $w_{s,s\pm 1}$. The trial move is accepted if the new site $s \pm 1$ is
empty, else it is rejected. A sweep of $n$ trial moves for a system having $n$ particles is considered as one Monte-Carlo step. We use periodic boundary condition. 
Note that in simulations we do not use the linearized versions of hopping rates,
unlike in theory with small potential strength $\l$. Instead we use the full non-linear forms. In all our simulations we keep $\l=0.5$, unlike the
perturbation theory where we assumed $\l \ll 1$. 
At the time-periodic steady state, the current is measured on each bond and then averaged over all bonds in the system, and  several time-periods. We also average over 
many initial conditions to obtain better statistics. For details of the parameter values used in simulations, see figure caption of Fig.~\ref{fig:Jro}.
All the simulation data show good agreement with predictions presented in Eq.(\ref{eq:JC1}).

For all the three variants of the model current vanishes as packing fraction $\r \to 0$ and $\r \to 1$ (close pack) limits. The first vanishing is due to absence of particles to carry current, and the second one is due to complete jamming. 
If all the lattice sites are occupied, within the discrete lattice random sequential dynamics, particles can not move. However, the detailed density dependence of current $\bar J$ shows three very different 
form for the three variants of the model considered. While for model A, $\bar J_A \sim \r^2 (1-\r)$, model B shows a dramatic effect of current reversal with changing density. 
For model B, the density dependence is $\bar J_B \sim \r(1-\r)(1-2\r)$, with a new zero in current appearing at the half filling $\r = 1/2$. This particular model has a  symmetry  under the exchange of
particles with holes together with swapping direction from right to left. Thus a phase factor $\phi$ that leads to free particle motion towards right, which is the dominant mode at low densities, will 
lead to {\em free} hole motion to right at high densities. Therefore, particle current changes direction from near $\r=0$ to near $\r=1$. At $\r=1/2$ the particle and hole currents cancel each other leading to $\bar J_B =0$. 
We  performed simulations for model C as well, and present the numerically obtained $\bar J_C$ in Fig.~\ref{fig:Jro}.
Our simulation results for model B and C agree well with theoretical predictions~(see  Fig.~\ref{fig:Jro}).  Note that, for model C, theory
predicts a density dependence $\bar J_C \sim \r (1-\r)^2$. 
In Fig.~\ref{fig:Jro} we have also plotted the theoretical prediction for model A, for comparison. 

\section{Outlook}
We presented a discrete pump model in which an external traveling wave potential leads to average directed motion of particles interacting via exclusion process. 
We discussed three possible choices of external potential dependent hopping rates, all of which obey the microscopic time-reversal symmetry. 
We studied how a resultant directed current depends on average density of particles. Using a perturbative expansion
for small strength of external potential with respect to thermal noise, we obtained analytic expressions for directed current, and compared our results with direct Monte-Carlo simulations to find good
agreement. While the time evolution of first order perturbation in local density and correlation functions, are independent of specific choice of the three variants of the lattice model discussed here, the expressions for directed current depend on the choice of local hopping rates.  
The dependence of average directed current on frequency and phase is the same across all the three choices of hopping rates [see Eq.(\ref{eq:JC})]. However, it is important to note that the detailed density dependence is  very different in the three choices of models (hopping rates) discussed -- while model A predicts $\bar J_A \sim \r^2(1-\r)$, model B and C predict $\bar J_B \sim \r(1-\r)(1-2\r)$ and $\bar J_C \sim \r (1-\r)^2 $ respectively [mean field limits of Eq.(\ref{eq:JC})]. The hopping rates chosen for model-A fails to generate any directed current in absence of particle exclusion -- density correlation turns out to be necessary. On the other hand, models B and C allows for driving of directed current for non-interacting particles, also. 
In Ref.~\cite{Chaudhuri2011} we argued that the model B is a natural choice, if one starts from the  corresponding Langevin equation and discretize its dynamics. This model shows a curious current reversal with increasing density of particles. 

However, later studies by us in continuum model showed very different density dependence~\cite{Chaudhuri2014}, in particular, absence of current reversal predicted by the discrete exclusion process in model-B. In order to discuss the continuum limit of our calculation, let us now use the lattice parameter $b$ explicitly. In the continuum limit, one has to take the lattice parameter $b/L \to 0$, with packing fraction $\r b \ll 1$. 
Since $b$ defines the length scale of inter-particle repulsion as well, in this limit, one obtains results valid for non-interacting continuum dynamics. However, in the real continuum system the space is continuum, but the hard core particles have finite size and repel each other. Thus the continuum limit of the discrete exclusion process used here, fails to capture the behavior of hard core particles moving in continuum space under stochastic thermal force, and time-oscillatory external potential. A correct discrete model would require the length scale of exclusion process to be defined as a new variable $\s=\nu b$, such that in the continuum limit, $b/L \to 0$ with $\nu \to \infty$ keeping $\s$ constant.  
Experimental realization of the model presented here looks possible, using colloidal particles confined in narrow channels driven by traveling wave potential. This would provide better insight into driven many-body dynamics, and could have potential applications.

%%%%%%%%%%%%%%%%%%%%%%%%%%%%%%%%%%%%%%%%%%%%%%%%%%%%%%%%%%%%%%%%%%%%%%
\ack
%%%%%%%%%%%%%%%%%%%%%%%%%%%%%%%%%%%%%%%%%%%%%%%%%%%%%%%%%%%%%%%%%%%%%% 
The author thanks Abhishek Dhar for numerous discussions, and a collaboration  on related topics which led to the publication of Ref.s\cite{Chaudhuri2011,Chaudhuri2014}.
%%%%%%%%%%%%%%%%%%%%%%%%%%%%%%%%%%%%%%%%%%%%
\vskip 2cm

\bibliographystyle{prsty}

\begin{thebibliography}{10}

\bibitem{Julicher1997}
F. J\"{u}licher, A. Ajdari, and J. Prost, Reviews of Modern Physics {\bf 69},
  1269  (1997).

\bibitem{Reimann2002}
P. Reimann, Physics Reports {\bf 361},  57  (2002).

\bibitem{Astumian2002}
R.~D. Astumian and P. H\"{a}nggi, Physics Today {\bf 55},  33  (2002).

\bibitem{Gadsby2009}
D.~C. Gadsby, A. Takeuchi, P. Artigas, and N. Reyes, Philosophical transactions
  of the Royal Society of London. Series B, Biological sciences {\bf 364},  229
   (2009).

\bibitem{Astumian2003}
R. Astumian, Physical Review Letters {\bf 91},  1  (2003).

\bibitem{Brouwer1998}
P. Brouwer, Physical Review B {\bf 58},  R10135  (1998).

\bibitem{Sela2006}
E. Sela and Y. Oreg, Physical Review Letters {\bf 96},  166802  (2006).

\bibitem{Cavaliere2009}
F. Cavaliere, M. Governale, and J. K\"{o}nig, Physical Review Letters {\bf
  103},  136801  (2009).

\bibitem{Strass2005}
M. Strass, P. H\"{a}nggi, and S. Kohler, Physical Review Letters {\bf 95},
  130601  (2005).
  
 \bibitem{Derenyi1995}
I. Der\'{e}nyi and T. Vicsek, Physical Review Letters {\bf 75},  374  (1995).

\bibitem{Derenyi1996}
I. Derenyi and A. Ajdari, Physical Review E {\bf 54},  R5  (1996).

\bibitem{Aghababaie1999}
Y. Aghababaie, G. Menon, and M. Plischke, Physical Review E {\bf 59},  2578
  (1999).

\bibitem{savel2004}
S. Savel’ev, F. Marchesoni, and F. Nori, Phys. Rev. E {\bf 70},  061107
  (2004). 

\bibitem{Jain2007}
K. Jain, R. Marathe, A. Chaudhuri, and A. Dhar, Physical Review Letters {\bf
  99},  190601  (2007).

\bibitem{Marathe2008}
R. Marathe, K. Jain, and A. Dhar, Journal of Statistical Mechanics: Theory and
  Experiment {\bf 2008},  P11014  (2008).

\bibitem{Chaudhuri2011}
D. Chaudhuri and A. Dhar, EPL (Europhysics Letters) {\bf 94},  30006  (2011).

\bibitem{Schutz2000}
G.~M. Sch{\"u}tz,  in {\em Phase transitions and Critical Phenomena}, edited by
  C. Domb and J. Lebowitz (Academic Press, London, 2000), pp.\ 3 -- 242.

\bibitem{Chaudhuri2014}
D. Chaudhuri, A. Raju, and A. Dhar, Phys. Rev. E {\bf 91}, 050103(R) (2015).

\end{thebibliography}

\end{document}